\documentstyle[prb,aps,epsfig,multicol]{revtex}

\begin{document}
\draft
\title{Temperature- and magnetic-field-dependent resistivity of MgB$_{2}$ sintered
at high temperature and high pressure condition }
\author{C. U. Jung, Min-Seok Park, W. N. Kang, Mun-Seog Kim, S. Y. Lee, and Sung-Ik
Lee\cite{email}}
\address{National Creative Research Initiative Center for\\
Superconductivity\\
and Department of Physics, Pohang University of Science and\\
Technology,\\
Pohang 790-784, Republic of Korea }
\date{Accepted by PhysicaC (2001).}
\maketitle

\begin{abstract}
We report the temperature- and magnetic-field-dependent resistivity of MgB$%
_{2}$ sintered at high temperature and high pressure condition. The
superconducting transition width for the resistivity measurement was about
0.4 K, and the low-field magnetization showed a sharp superconducting
transition with a transition width of about 1 K. The resistivity in the
normal state roughly followed $T^{2}$ behavior with smaller residual
resistivity ratio (RRR) of 3 over broad temperature region above 100 K
rather than reported $T^{3}$ behavior with larger RRR value of $\sim 20$ in
the samples made at lower pressures. Also, the resistivity did not change
appreciably with the applied magnetic field, which was different from
previous report. These differences were discussed with the microscopic and
structural change due to the high-pressure sintering.
\end{abstract}


\begin{multicols}{2}
\section{Introduction}

Very recently, superconductivity at about 40 K was discovered in MgB$_{2}$. 
\cite{Akimitsu} Conventional BCS superconductivity has been proposed for
this compound and due to the enhanced phonon frequency from the light ionic
masses, the superconducting transition temperature is expected to be high. 
\cite{Kortus} A shift in the $T_{c}$ due to the boron isotope has been
reported with an isotope critical exponent of $\alpha _{B}\sim 0.26$.\cite
{Budko} In addition, several thermodynamic parameters have been measured in
this compound and these include a upper critical field $H_{c2}$, its slope $%
dH_{c2}/dT,$ the Ginzburg-Landau parameter $\kappa $, zero-temperature
coherence length $\xi (0)$, and penetration depth $\lambda (0)$ etc.\cite
{Finnemore} It is quite interesting that these thermodynamic parameters are
nearly the same as those for Sr$_{0.9}$La$_{0.1}$CuO$_{2}$,\cite{MSKim3dIL}
whose infinite-layer structure consisting of a conducting plane (CuO$_{2}$)
and a\ metallic spacer layer (Sr,La) is quite similar to the structure of MgB%
$_{2}$.

The type of carrier was predicted to be positive with boron planes acting
like the CuO$_{2}$ planes in cuprate high-temperature superconductors\cite
{Hirsch}, which was confirmed by a Hall measurement by us.\cite{WNKang} With
high carrier density (%
\mbox{$>$}%
10$^{23}$ cm$^{-3}$) revealed by the Hall measurement, the material have
been reported to have characteristic metallic transport behaviors. The
transport measurements for MgB$_{2}$ synthesized at lower pressures ($p\ll 1$
GPa) showed that a residual resistivity ratio between 300 K and 40 K was
more than 20 with room temperature resistivity of about 10 $\mu \Omega $cm,
and the resistivity value increases by several tens of percent in the
magnetic field of 5 Tesla, and overall temperature dependence followed a $%
T^{3}$ behavior in the normal state.\cite{Finnemore} However, samples made
at lower pressures were reported to be rather porous and mechanically weak, 
\cite{Paul} thus the concrete transport properties should be established for
compact form of samples especially for device application.

In this paper we report the physical properties of hard and dense MgB$_{2}$
sintered at high temperature and high pressure ($p\sim 3$ GPa). This sample
was strong enough to prepare an optically clean surface for the reflectivity
measurement by polishing. We found that $T_{c}$ onset decreased by about 0.5
K due to the high-pressure sintering and overall temperature dependence of
the resistivity in the normal state followed a $T^{2}$ and residual
resistivity ratio was less than 3. And near $T_{c}$, the change of
resistivity with magnetic field up to 5 Tesla All these resistivity
behaviors were different from those for samples made at lower pressures.\cite
{Finnemore} These differences were discussed with the microscopic and
structural changes caused by the high-pressure sintering.

\section{Experimental}

High-pressure sintering was performed with a 12-mm cubic multi-anvil-type
press.\cite{JungLa} Commercially available powder of MgB$_{2}$ (Alfa Aesar)
was used to make pellets.\cite{Alfa} The pellets were put into a Au capsule
in a high-pressure cell. One group of pellets were pressurized upto 3 GPa
without subsequent heat treatment, to make a `cold-pressed'
sample(CP-sample) for resistivity measurement and the other group of the
pellets was heated after pressurization to make a `hot-pressed'
sample(HP-sample). A {\it D}-type thermocouple was inserted near the Au
capsule to monitor the temperature. It took about 2 hours to pressurize the
cell to 3 GPa. After the pressurization, the heating power was increased
linearly and then maintained constant for 2 hours. The sample was sintered
at a temperature of $850\sim 950$$^{\circ }$C and then quenched to room
temperature. The weight of the sample obtained in one batch was about 130
mg, and the size was about 4.5 mm in diameter and 3.3 mm in height.

A SQUID magnetometer (Quantum Design, MPMS{\it XL}) was used to measure the
low-field magnetization of the samples. A scanning electron microscope (SEM)
was used to investigate the surface morphology. For the resistance
measurement, we cut the HP-sample by using a diamond saw with a coolant and
then polished it into a rectangular solid shape with dimension. The
resistance curve, $R$($T$), was measured using the standard 4-probe
technique.\cite{WNKang}

\section{Data and discussion}

\begin{figure}[tb]
\begin{center}
\epsfig{file=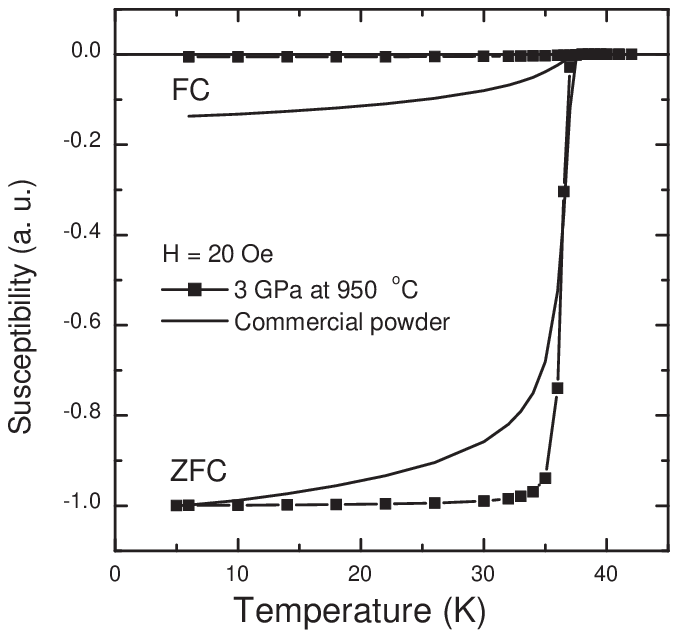,width=0.55\textwidth,height=0.4\textheight,clip=
,angle=0}
\end{center}
\caption{Normalized magnetic susceptibility from the low-field magnetization 
$M(T)$\ of MgB$_{2}$. $M(T)$ for zero-field-cooling and field-cooling states
were measured at 20 Oe. The curve with a broad transition was for the
as-purchased powder (50 mg), and the curve with a sharp transition was for
the HP-sample (120 mg).}
\label{lowfieldMT}
\end{figure}

Figure \ref{lowfieldMT} shows the normalized magnetic susceptibility, $4\pi
\chi (T)$, from the measured\ low-field magnetization data for two kinds of
MgB$_{2}$. The curve with the broad transition is for as-purchased powder,
the other curve with the symbols is for the HP-sample. The magnetic
susceptibility data show that the superconducting transition width decreased
from about 10 K to 1 K af ter the high-pressure sintering. The decreased
field-cooling signal in the magnetic susceptibility for HP-sample indicates
that the flux pinning is greatly enhanced and suggests a higher possibility
of high current superconducting applications in the compact bulk form, which
was also verified from the bulk-pinning behaviors in the magnetic hysteresis 
$M(H)$.\cite{MSKim} The transition temperature of the HP-sample was about
37.5 K, was slightly lower than 38 K for the as-purchased powder.

\begin{figure}[tbp]
\begin{center}
\epsfig{file=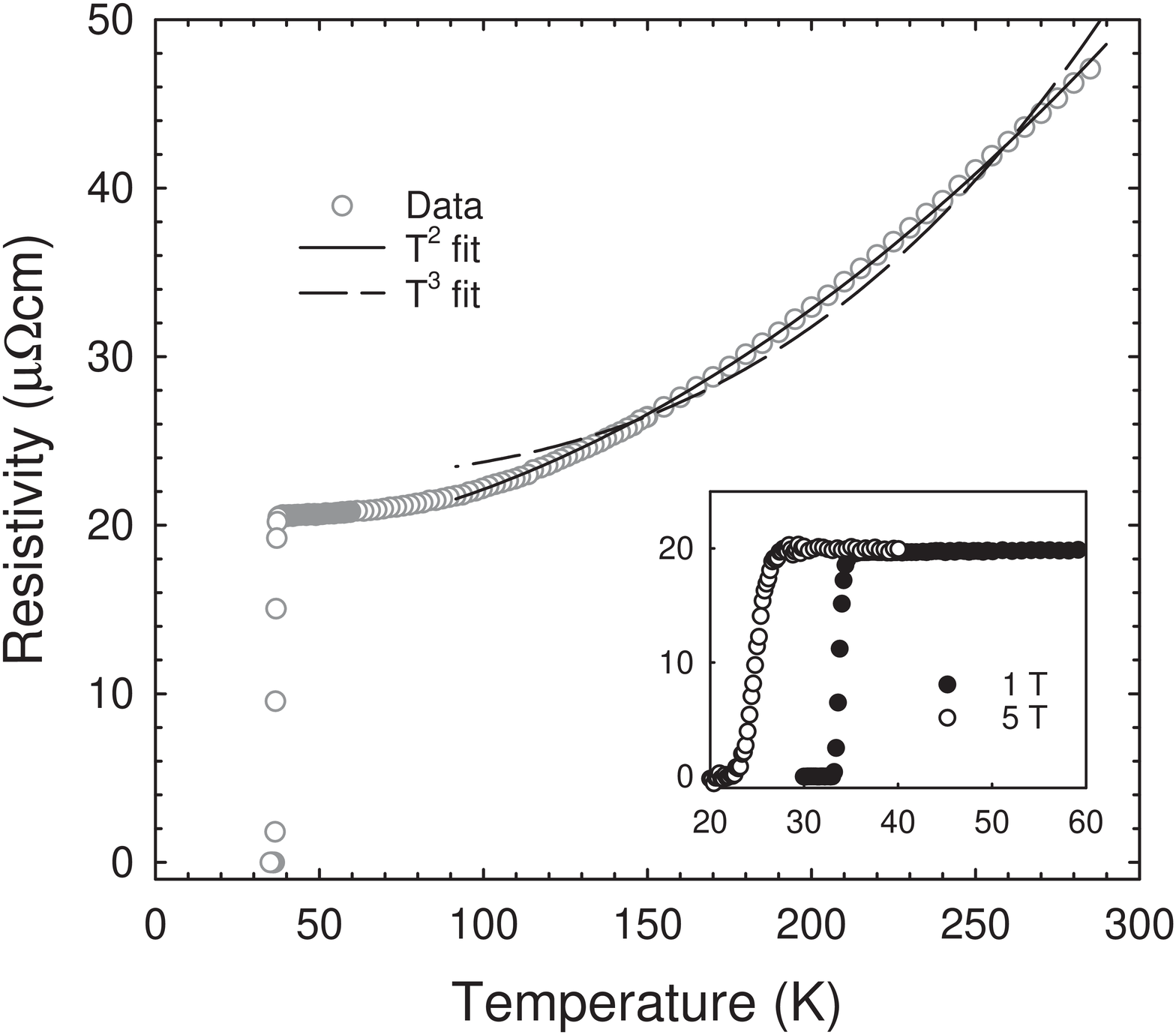,width=0.55\textwidth,height=0.4\textheight,clip=
,angle=0}
\end{center}
\caption{Resistivity of HP-sample of MgB$_{2}$. Symbols denote the data and
the two lines are fitting lines. Above $T_{c}$, the temperature dependence
of the resistance followed a $T^{2}$ behavior (solid line through the data)
rather than reported previously $T^{3}$ behavior (dashed line). The
transition width was about 0.4 K.\ The inset shows the dependence of the
resistivity on the external magnetic field. This dependence is different
from the previous results that the resistivity near $T_{c}$ doubles upon
applying high magnetic field $H=9$ T. }
\label{RT}
\end{figure}

Figure \ref{RT} shows the temperature- and the magnetic-field dependences of
the resistivity. The transition width was about 0.4 K for a 10 to 90\% drop
of the resistivity curve. The resistivity value was $\rho \sim $ 50 and 21 $%
\mu \Omega $cm at room temperature and at $T=40$ K. The metallic nature of
HP-sample can be also easily inferred from its shiny surface with dark
yellow tint. The overall temperature dependence of the resistivity follows a 
$a+bT^{2}$ behavior rather than the $a+bT^{3}$ form previously reported. 
\cite{Finnemore} The solid fitting line of the $a+bT^{2}$ form in Fig. \ref
{RT} nearly overlays the data, but the dashed line of the\ $a+bT^{3}$ form
shows a significant deviation from the data. The parameters for the $%
a+bT^{2} $ fitting are $a=2.000\times 10^{-5}$ $\mu \Omega $cm and\ \ $%
b=3.566\times 10^{-10}$ $\mu \Omega $cm/K$^{2}$. A fitting at lower
temperature region below 100 K was not sensitive enough to distinguish
clearly either form. The inset of Fig. \ref{RT} shows the resistivity
measured under an external magnetic field. The resistance in the normal
state does not change appreciably on the external magnetic field. Thus, the
dependence of the resistivity either on the temperature or on the external
magnetic field was quite different from the previous results measured for
samples made at lower pressures.\cite{Finnemore}

\begin{figure}[tbp]
\begin{center}
\epsfig{file=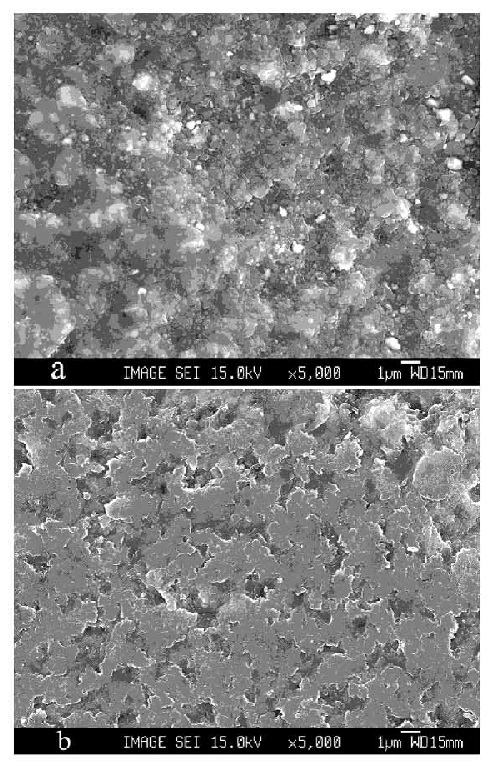,width=0.55\textwidth,height=0.55\textheight,clip
=,angle=0}
\end{center}
\caption{SEM\ pictures of (a) the CP-sample and (b) the HP-sample which is a
heat treated MgB$_{2}$ at 3 GPa. The scale bars indicate 1 $\protect\mu $m
for both pictures.}
\label{SEM}
\end{figure}

Figure \ref{SEM} shows SEM pictures for both samples. In Fig. \ref{SEM}.
(a), the grain size of the CP-sample is much less than $1$ $\mu $m. The
grains in the HP-sample are well connected as shown in Fig. \ref{SEM}. (b).
We cannot even distinguish the grain boundaries over wide regions. The
microscopic connections between the grains may be the reasons for HP-sample
being strong and dense macroscopically.

The above resistivity behaviors for HP-samples sintered at high-pressure $%
p\gg 1$ GPa were also observed by other group,\cite{Takano} thus might be
somehow intrinsic. Most probable extrinsic origin for the different
resistivity behaviors for samples made at higher pressures would be the
appearance of inter-grain impurities during the high-pressure sintering
which wrap around the grains and block the inter-grain current transport.
However our recent study using a high-resolution transmission electron
microscope showed that the impurities were well isolated from the major MgB$%
_{2}$ phases, not forming the inter-grain layer.\cite{Sung} Then the strong
connectivity of the grains without inter-grain impurity might suggest that
intra-grain contribution itself dominates the different resistivity
behavior. The lattice parameters directly obtained from high resolution TEM
images were the nearly same as those of lower-pressure samples measured near
3 GPa.\cite{Sung,cond-mat/0102480,cond-mat/0102507} Thus the strain caused
by the high-pressure seems to remain appreciably after the sintering, which
could answer partly the different transport behavior in some way. The
observed difference of $T_{c}$ between HP-sample and CP-sample was less than
about 0.5 K, which is much smaller than reported $T_{c}$ reduction by 4.8 K
upon applying 3\ GPa.\cite{Chu} The different $T_{c}$ reduction may be
partly due to different pressurization method and/or the different
distribution of strains inside the sample. The identification of the exact
origin for the different resistivity behaviors still needs more studies.

\section{summary}

In summary, we report the temperature- and magnetic-field-dependant
resistivity of a hard and dense MgB$_{2}$ sintered at high temperature and
high pressure ($p\sim 3$ GPa). The superconducting transition width for the
resistivity measurement was about 0.4 K, and the resistivity in the normal
state followed a behavior of $T^{2}$. The absolute values of the resistivity
at room temperature and at just above $T_{c}$ were 50 to 21 $\mu \Omega $cm
respectively. Also, the resistivity in the normal state did not change
appreciably with the applied magnetic field up to 5 T. These behaviors are
different from those for samples made at lower pressures and maybe are
partly due to the presence of strain caused by high-pressure condition for
the sintering.

\acknowledgments
This work is supported by the Ministry of Science and Technology of Korea
through the Creative Research Initiative Program.

\newpage
\end{multicols}

\newpage


\begin{references}
\bibitem[*]{email}  Electronic address: silee@postech.ac.kr

\bibitem{Akimitsu}  J. Nagamatsu, N. Nakagawa, T. Muranaka, Y. Zenitani, and
J. Akimitsu, Nature {\bf 410}, 63 (2001).

\bibitem{Kortus}  J. Kortus, I. I. Mazin, K. D. Belashchenko, V. P.
Antropov, and L. L. Boyer, cond-mat/0101446 (2001).

\bibitem{Budko}  S. L. Bud'ko, G. Lapertot, C. Petrovic, C. E. Cunningham,
N. Anderson, and P. C. Canfield, Phys. Rev. Lett. {\bf 86}, 1877 (2001).

\bibitem{Finnemore}  D. K. Finnemore, J. E. Ostenson, S. L. Bud'ko, G.
Lapertot, and P. C. Canfield, Phys. Rev. Lett. {\bf 86}, 2420 (2001).

\bibitem{MSKim3dIL}  Mun-Seog Kim, C. U. Jung, J. Y. Kim, Jae-Hyuk Choi, and
Sung-Ik Lee, cond-mat/0102420, (2001).

\bibitem{Hirsch}  J. E. Hirsch, cond-mat/0102115 (2001).

\bibitem{WNKang}  W. N. Kang, C. U. Jung, Kijoon H. P. Kim, Min-Seok Park,
S. Y. Lee, Hyeong-Jin Kim, Eun-Mi Choi, Kyung Hee Kim, Mun-Seog Kim, and
Sung-Ik Lee, cond-mat/0102313 (2001).

\bibitem{Paul}  Mackenie Paul, private communication.

\bibitem{JungLa}  C. U. Jung, J. Y. Kim, Mun-Seog Kim, Min-Seok Park,
Heon-Jung Kim, Yushu Yao, S. Y. Lee, and Sung-Ik Lee, (submitted to Physica
C).

\bibitem{Alfa}  Alfa Aesar, A Johnson Matthey Company,\ Stock \# 88149:
magnesium boride, 98 \% (assay) MgB2 (possible impurities are not specified).

\bibitem{MSKim}  Mun-Seog Kim, C. U. Jung, Min-Seok Park, S. Y. Lee, Kijoon
H. P. Kim, W N. Kang, and Sung-Ik Lee, cond-mat/0102338 (2001).

\bibitem{Takano}  Y. Takano, H. Takeya, H. Fujii, H. Kumakura, T. Hatano, K.
Togano, H. Kito, and H. Ihara, cond-mat/0102167 (2001), D. D. Lawrie, J. P.
Franck, and Granwen Zhang, Post deadline oral session on MgB2 during APS
2001 march meeting.

\bibitem{Sung}  Gun Yong Sung, Sang Hyeob Kim, JunHo Kim, Dong Chul Yoo, Ju
Wook Lee, Jeong Yong Lee, C. U. Jung, Min-Seok Park, W. N. Kang, Du
Zhonglian, and Sung-Ik Lee, cond-mat/0102498, (2001).

\bibitem{cond-mat/0102480}  T. Vogt, G. Schneider, J. A. Hriljac, G. Yang,
and J. S. Abell, cond-mat/0102480 (2001).

\bibitem{cond-mat/0102507}  K. Prassides, Y. Iwasa, T. Ito, D. H. Chi, K.
Uehara, E. Nishibori, M. Takata, S. Sakata, Y. Ohishi, O. Shimomura, T.
Muranaka, and J. Akimitsu, cond-mat/0102507 (2001).

\bibitem{Chu}  B. Lorenz, R. L. Meng, and C. W. Chu, cond-mat/0102264 (2001).
\end{references}
\end{document}